\documentclass[preprint,onecolumn,nofootinbib]{revtex4-1}
\pdfoutput=1
\usepackage[colorlinks=true,linkcolor=blue,urlcolor=blue,filecolor=black,citecolor=red,pdfstartview=FitV,pdftitle={},pdfsubject={},pdfkeywords={},pdfpagemode=None,bookmarksopen=true]{hyperref}
\usepackage{graphicx}%Include figure files
\usepackage{amsmath,tikz}
\usepackage{amsfonts}
\usepackage{amssymb,ulem}
\usepackage{color}%
\usepackage{CJK}
\usepackage{dcolumn}% Align table columns on decimal pointl
\newcommand{\f}{\begin{equation}}
\newcommand{\ff}{\end{equation}}
\newcommand{\fa}{\begin{eqnarray}}
\newcommand{\ffa}{\end{eqnarray}}

\begin{document}
\title{Mixed State Entanglement for Holographic Axion Model}
\author{Yi-fei Huang $^{1}$}
\author{Zi-jian Shi $^{1}$}
\author{Chao Niu $^{1}$}
\author{Cheng-yong Zhang $^{1}$}
\author{Peng Liu $^{1}$}
\thanks{jnuhyf@stu2017.jnu.edu.cn}
\thanks{szj@stu2016.jnu.edu.cn}
\thanks{niuchaophy@gmail.com}
\thanks{zhangcy@email.jnu.edu.cn}
\thanks{phylp@jnu.edu.cn}
\affiliation{
	$^1$ Department of Physics and Siyuan Laboratory, Jinan University, Guangzhou 510632, China}
\begin{abstract}

	We study the mixed state entanglement in holographic axion model. We find that the holographic entanglement entropy (HEE), mutual information (MI) and entanglement of purification (EoP) exhibit very distinct behaviors with system parameters. The HEE exhibits universal monotonic behavior with system parameters, while the behaviors of MI and EoP relate to the specific system parameters and configurations. We find that MI and EoP can characterize mixed state entanglement better than HEE, since they are less affected by thermal effects.
	Moreover, we argue that EoP is more suitable for describing mixed state entanglement than MI. Because the MI of large configurations are still dictated by the thermal entropy, while the EoP will never be controlled only by the thermal effects.

\end{abstract}
\maketitle
\tableofcontents
\section{Introduction}\label{mi-inho}

Quantum entanglement is one of the most striking features of quantum systems that distinguish them from classical systems. Recently, quantum entanglement has been extensively studied in quantum information, condensed matter theory, quantum gravity, and become the core of interdisciplinary of these fields. Quantum entanglement can characterize the quantum phase transition of strong correlation systems and the topological quantum phase transitions, and plays a key role in the emergence of spacetime \cite{Osterloh:2002na,Amico:2007ag,Wen:2006topo,Kitaev:2006topo,Ryu:2006bv,Hubeny:2007xt,Lewkowycz:2013nqa,Dong:2016hjy}.

There are many measures of quantum entanglement, such as entanglement entropy (EE), R\'enyi entanglement entropy, negativity, mutual information (MI), and so on. Each metric can characterize different aspects of quantum entanglement. EE is a good measure for pure state entanglement, but is unsuitable for mixed state entanglement. In fact, mixed state are more common than pure states. For example, any system with finite temperature corresponds to a mixed state, while only zero temperature system can be a pure state. Therefore, many new entanglement metrics have been proposed to characterize mixed state entanglement, such as the non-negativity, entanglement of purification and the entanglement of formation \cite{vidal:2002,Horodecki:2009review}. However, entanglement measures are notoriously difficult to calculate, because the Hilbert space of quantum systems is often extremely large.

Gauge/gravity duality is a powerful tool for studying strongly correlated systems, and it also brings geometric prescriptions for entanglement related physical quantities. The holographic entanglement entropy (HEE) associates the EE of a subregion with the area of the minimum surface in the dual gravity system \cite{Ryu:2006bv}. HEE has many important applications, such as characterizing quantum phase transitions and thermodynamic phase transitions \cite{Nishioka:2006gr,Klebanov:2007ws,Pakman:2008ui,Zhang:2016rcm,Zeng:2016fsb}. The HEE proposal opened the door for exploring the information-related properties in holographic theories. For instance, as a more general measure of entanglement, R\'enyi entropy is proposed to be proportional to the minimal area of cosmic branes \cite{Dong:2016fnf}. Moreover, the butterfly effect as a typical phenomenon of quantum chaos has been extensively studied in holographic theory recently \cite{Shenker:2013pqa,Sekino:2008he,Maldacena:2015waa,Donos:2012js,Blake:2016wvh,Blake:2016sud,Ling:2016ibq,Ling:2016wuy,Wu:2017mdl,Liu:2019npm}. In addition, holographic duality of quantum complexity has also been proposed \cite{Brown:2015lvg,Brown:2015bva,Chapman:2016hwi,Ling:2018xpc,Chen:2018mcc,Yang:2019gce,Ling:2019ien}. Recently, the purification of purification (EoP) was associated with the area of the minimum cross-section of the entanglement wedge \cite{Takayanagi:2017knl,Nguyen:2017yqw}. The geometric dual of EoP provides a novel tool for studying the mixed state entanglement.

At present, HEE has been widely studied, but the research on mixed state entanglement - MI and EoP, is still to be enhanced. In particular, the difference between MI and EoP, and their effectiveness in characterizing mixed state entanglement, are still unclear. For this purpose, we study the properties of HEE, MI and EoP, in a holographic axion model. We choose axion model because it is simple enough to have analytical solutions, meanwhile it has some important properties such as momentum dissipation.

We organize this paper as follows: we introduce the holographic axion model in Sec. \ref{subsec:model}, entanglement measures (HEE, MI, EoP) and their holographic duality in Sec. \ref{subsec:info}. In Sec. \ref{mi-intro}, we discuss the properties of these three information related physical quantities in axion model. Finally, we summarize in Sec. \ref{sec:discuss}.

\section{The holographic axion model and information related quantities}\label{sec:model_info}

In this section, we introduce the holographic axion model, as well as the concepts and calculations of HEE, MI and EoP.

\subsection{Holographic axion model}\label{subsec:model}

We consider the following action \cite{Andrade:2013gsa,Gouteraux:2016wxj},
\begin{equation}\label{eq:action}
	S = \int d ^ { 4 } x \sqrt { - g } \left[ R + 6 - V ( X ) - \frac { 1 } { 4 } F _ { \mu \nu } F ^ { \mu \nu }\right],
\end{equation}
where $V(X)$ is the kinematic term for the axion fields. The ansatz and the corresponding solutions are,
\begin{equation}
	\begin{aligned}
		d s ^ { 2 } & = - f ( r ) d t ^ { 2 } + \frac { 1 } { f( r ) } d r ^ { 2 } + r^2 \delta_{ij}{dx}^i{dx}^j  \\
		A _ { t }   & = \mu - \frac { \rho } { r },\quad \psi ^ { I } = k \delta _ { i } ^ { I } x ^ { i }, \quad X = \frac{1}{2}g^{ab} \delta_{IJ} \partial_a \psi^I \partial_b \psi^J, \quad \text{where}\;i,j,I,J = 1,2,
	\end{aligned}
\end{equation}
with
\begin{equation}\label{eq:fexp}
	f( r ) =  \frac { \rho ^ { 2 }} { 4 r } \left( \frac { 1 } { r } - \frac { 1 } { r _ { h } } \right) + \frac { 1 } { 2 r } \int _ { r _ { h } } ^ { r } \left[ 6 u ^ { 2 } - V \left( \frac { k ^ { 2 } } { u ^ { 2 } } \right) u ^ { 2 } \right] d u.
\end{equation}
The $\psi^I$ represents the linear axion field with a constant linear factor $k$. The $A$ is the Maxwell field, and $ \mu,\,\rho$ represent the chemical potential and the charge density of the dual field theory. The regularity of the Maxwell field on the horizon requires that $A_t|_{r_h} = 0$, \emph{i.e.}, $\rho =\mu r_h$. The linear axion fields break the translational symmetry, so the system has finite DC conductivity, which reads \cite{Gouteraux:2016wxj},
\begin{equation}\label{dc-cond}
	\sigma _ { D C } = 1 +  \frac { \mu ^ { 2 } } { k ^ { 2 } \tilde V \left( r _ { h } \right) },
\end{equation}
with $\tilde { V } ( r ) \equiv \sum _ { n = 1 } ^ { \infty } 2n V' \left( X^n /2 \right) ^ { n - 1 }$.

For clarity, we focus on the $V(X) = X$ case, which goes back to \cite{Andrade:2013gsa}.\footnote{We have also checked the $V(X) = X^2$ case, and obtained qualitatively the same results as the $V(X) = X$ case.}
The system is invariant under the rescaling,
\begin{equation}\label{eq:resclaing}
	(t,x,y) \to \alpha (t,x,y),\, (r,k,\mu)\to (r,k,\mu)/\alpha,\, (f(r),\rho) \to (f(r),\rho)/\alpha^2.
\end{equation}
We focus on scaling-invariant physical quantities, so we adopt $\mu$ as the scaling unit by setting $\mu = 1$. Hawking temperature is given by,
\begin{equation}\label{eq:hawkingt}
	T =\left.\frac{1}{4\pi} \frac{df}{dr}\right|_{r=r_{h}} =-\frac{r_h^2+2 k^4-12 r_h^4}{16 \pi  r_h^3}.
\end{equation}

\subsection{The holographic information-related quantities}\label{subsec:info}

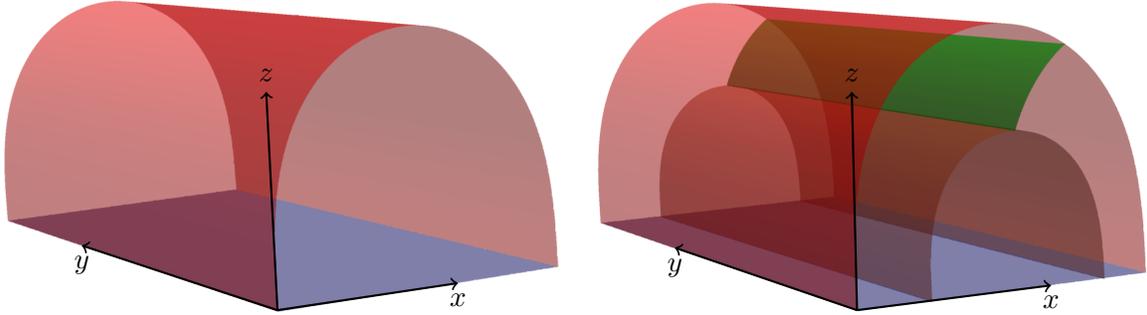
\begin{figure}
	\begin{tikzpicture}[scale=1]
		\node [above right] at (0,0) {\includegraphics[width=7.5cm]{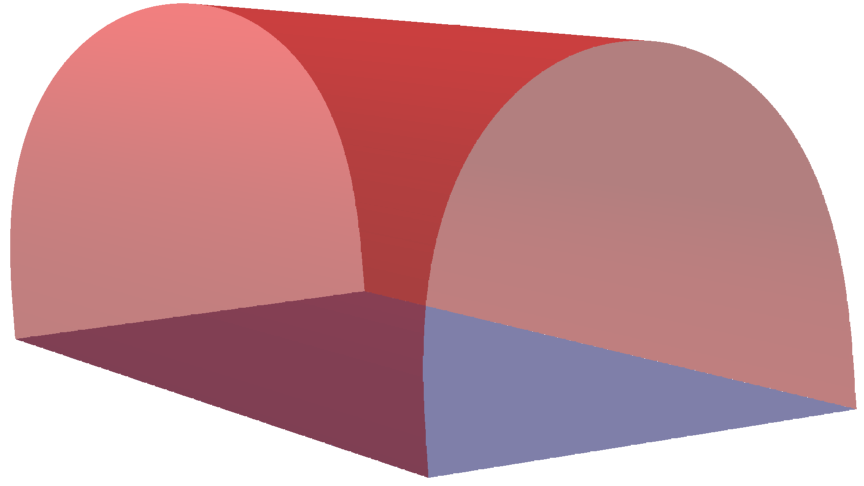}};
		\draw [right,->,thick] (3.85, 0.22) -- (6.25, 0.58) node[below] {$x$};
		\draw [right,->,thick] (3.85, 0.22) -- (1.25, 1.08) node[below] {$y$};
		\draw [right,->,thick] (3.85, 0.22) -- (3.7, 3.125) node[above] {$z$};
	\end{tikzpicture}
	\begin{tikzpicture}[scale=1]
		\node [above right] at (0,0) {\includegraphics[width=7.5cm]{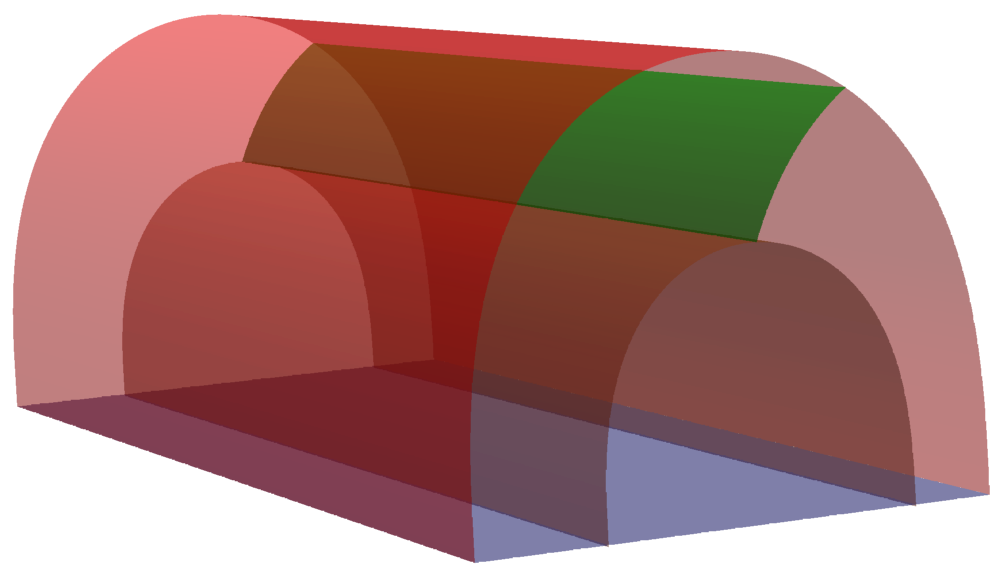}};
		\draw [right,->,thick] (3.67, 0.22) -- (6.25, 0.55) node[below] {$x$};
		\draw [right,->,thick] (3.67, 0.22) -- (1.25, 1.05) node[below] {$y$};
		\draw [right,->,thick] (3.67, 0.22) -- (3.6, 3.125) node[above] {$z$};
	\end{tikzpicture}
	\caption{The left plot: The cartoon of a minimum surface. The right plot: The cartoon of the minimum cross-section (green surface) of the entanglement wedge.}
	\label{fig:msd1}
\end{figure}

The HEE is identified as the area of the minimum surface stretching into the bulk. It is more convenient to use $ z \equiv \frac{r_{h}}{r} $ coordinate for numerical calculation, where
\begin{equation}
	ds^2=\frac{r^{2}_{h}}{z^{2}} \left[ - g(z) dt^{2} + \frac{dz^{2}}{g(z)r_{h}^{2}} + dx^{2} + dy^{2} \right],
\end{equation}
with
\begin{equation}\label{eq:ftog}
	g(z) = \frac{(z-1) \left(z^3 r_h^2-4 \left(z^2+z+1\right) r_h^4+2 k^4 z^3\right)}{4 r_h^4}.
\end{equation}
For simplicity, we consider the infinite strip configuration along $y$-axis (see the left plot of Fig. \ref{fig:msd1}). The minimum surface is invariant along $y$-axis, and hence can be described by $z(x)$. Therefore, solving the minimum surface only involves in ordinary differential equations, instead of partial differential equations. The HEE $S$ of the minimum surface and the corresponding width $w$ of the strip are,
\begin{equation}\label{eq:heeawidth}
	\begin{aligned}
		S\left(z_*\right) & =\int_\epsilon^{z_*} \frac{4 z_*^2 r_h^3}{z^2} {\mathcal T}^{-1/2} \, dz, \\
		w\left(z_*\right) & =\int_0^{z_*} 4 z^2 r_h \mathcal{T}^{-1/2} \, dz.
	\end{aligned}
\end{equation}
where $\mathcal T \equiv (1-z) \left(z^4-z_*^4\right) \left(z^3 r_h^2-4 \left(z^2+z+1\right) r_h^4+2 k^4 z^3\right)$ , and $z_*$ represents the top of the minimum surface. The asymptotic AdS boundary leads to a divergent HEE, therefore we introduce a cutoff $\epsilon$. Subtracting a cutoff-dependent quantity, the regularized HEE is given by,
\begin{equation}\label{eq:hee}
	S\left(z_*\right) =\int_0^{z_*} \left(
	\frac{4 z_*^2 r_h^3}{z^2} {\mathcal T}^{-1/2}-\frac{2r_h}{z^2}
	\right) \, dz-\frac{2r_h}{z_*}.
\end{equation}

\begin{figure}[h]
	\centering
	\includegraphics[width = 0.5 \textwidth]{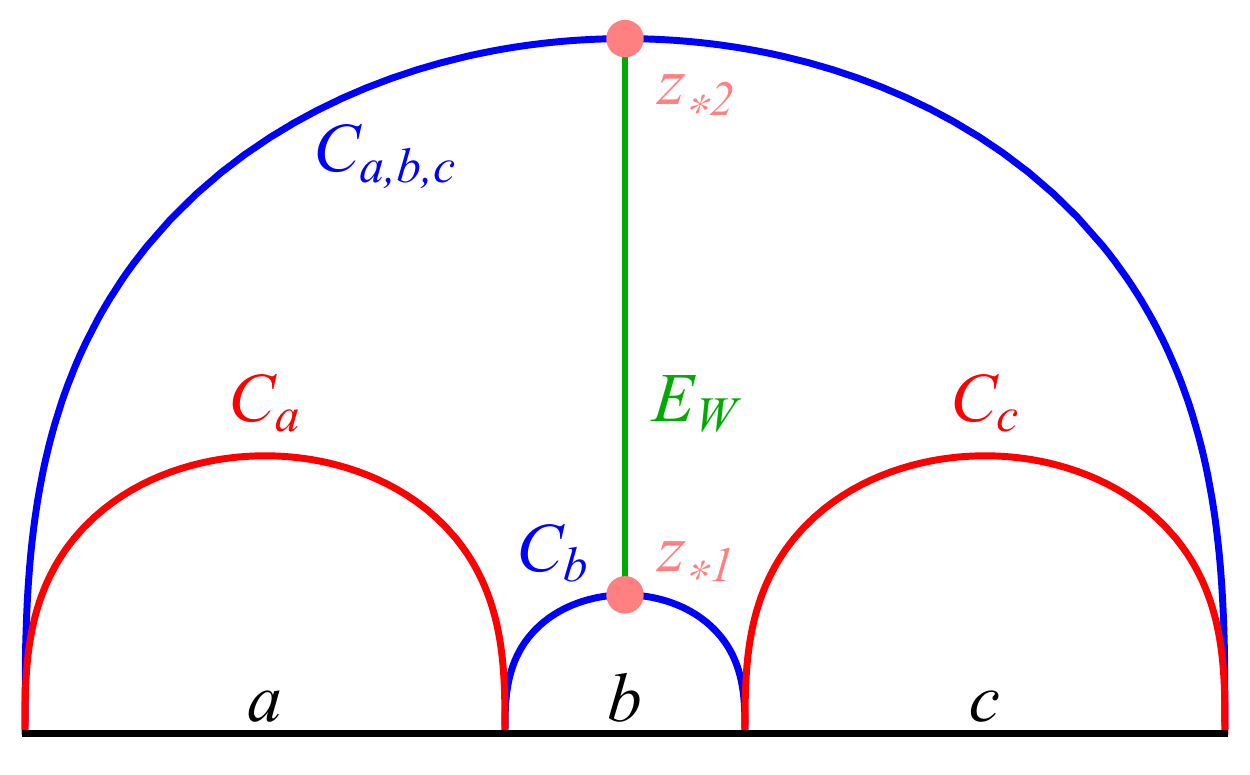}
	\caption{A cartoon of the mutual information and EoP for symmetric configurations. A non-trivial MI equals the area of the blue surfaces (connected configuration) minus the area of the red curves (disconnected configuration). The vertical green line represents the minimal cross section, that connects the tops ($z_{*1}$ and $z_{*2}$) of $C_b$ and $C_{a,b,c}$.}
	\label{fig:midemo}
\end{figure}

The mutual information between two separate subsystems $A$ and $C$ (with separation $B$) is defined by the following formula,
\begin{equation} \label{eq:eps}
	I(A;C)=S(A)+S(C)-S(A\cup C).
\end{equation}
Apparently, a non-trivial MI requires $S(A\cup C) = S(B) + S(A\cup B\cup C)$ (see
Fig. \ref{fig:midemo} for cartoon).
MI measures the entanglement between subsystems. Unlike the holographic entanglement entropy, this definition cancels out the divergence from the AdS. Moreover, MI also partly cancels out the thermal entropy contribution \cite{Fischler:2012uv}.
For convenience, let's consider parallel infinite stripes $A$ and $C$ with separation $B$, whose widths are $a,\,b$ and $c$ respectively. Therefore we can label a configuration with $(a, b, c)$ (see the right plot of Fig. \ref{fig:msd1}).

The entanglement of purification, defined as the minimum entanglement entropy among all possible purification, is a distinct entanglement measure from the HEE and MI \cite{Terhal:2002}. Recently, it is proposed that the EoP is proportional to the minimal cross-section for connected configuration of MI (see the right plot of \ref{fig:msd1}) \cite{Takayanagi:2017knl},
\begin{equation}\label{heop:def}
	E_{W}\left(\rho_{AC}\right) = \min_{\Sigma_{AC}} \left( \frac{\text{Area} \left(\Sigma_{AC}\right)}{4G_{N}}\right).
\end{equation}
EoP can measure mixed state entanglement for separate subsystems, and satisfies several important inequalities \cite{Terhal:2002,Takayanagi:2017knl}.


Next, we explore the HEE, MI and EoP on the axion model.

\section{The entanglement measures on the axion model}\label{mi-intro}

\subsection{Holographic Entanglement Entropy}\label{subsec:hee}

\begin{figure}[htbp]
	\centering
	\includegraphics[width=0.75\textwidth]{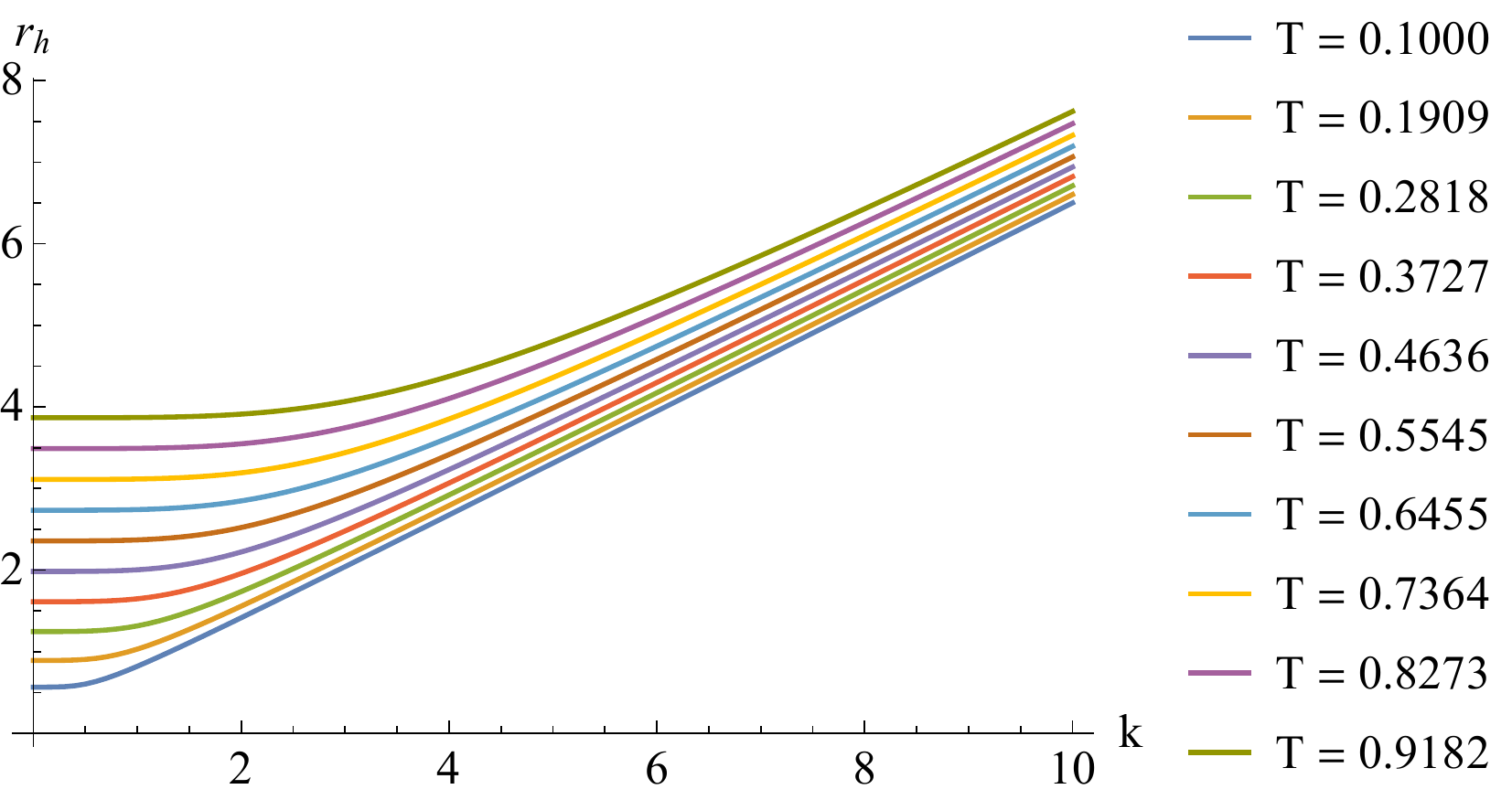}
	\caption{The $ r_{h} $ vs $ k $.}
	\label{fig:rhvsk4t}
\end{figure}

The Hawking temperature \eqref{eq:hawkingt} suggests that $r_h$ is a function of $(k,T)$, which we plot in Fig. \ref{fig:rhvsk4t}. At first, $r_h$ hardly changes with $k$ for small $k$. With the increase of $k$, $r_h$ gradually increases with $k$, and finally exhibits a linear relationship with an constant slope. This fact can be deduced from \eqref{eq:hawkingt}. At a fixed temperature $T$, we obtain
\begin{equation}\label{eq:drhdk}
	\left.\frac{\partial r_h}{\partial k}\right|_T = \frac{8 k^3 r_h}{12 r_h^4+r_h^2+6 k^4}.
\end{equation}
Solving $k$ from \eqref{eq:hawkingt} and inserting it into \eqref{eq:drhdk} we obtain that,
\begin{equation}\label{eq:drhdklimit}
	\lim_{r_h\to\infty} \left.\frac{\partial r_h}{\partial k}\right|_T = \lim_{r_h\to\infty} \frac{2 \sqrt[4]{2} \sqrt{r_h} \left(-16 \pi  T r_h+12 r_h^2-1\right){}^{3/4}}{-24 \pi  T r_h+24 r_h^2-1} = 6^{-1/4}.
\end{equation}
Therefore, the slope of $r_h$ vs $k$ at any temperature approaches $6^{-1/4}$ for large $k$. The linear relationship between $r_h$ and $k$ in large $k$ limit plays an important role in the behavior of information related physical quantities.

\begin{figure}[htbp]
	\centering
	\includegraphics[width=0.75\textwidth]{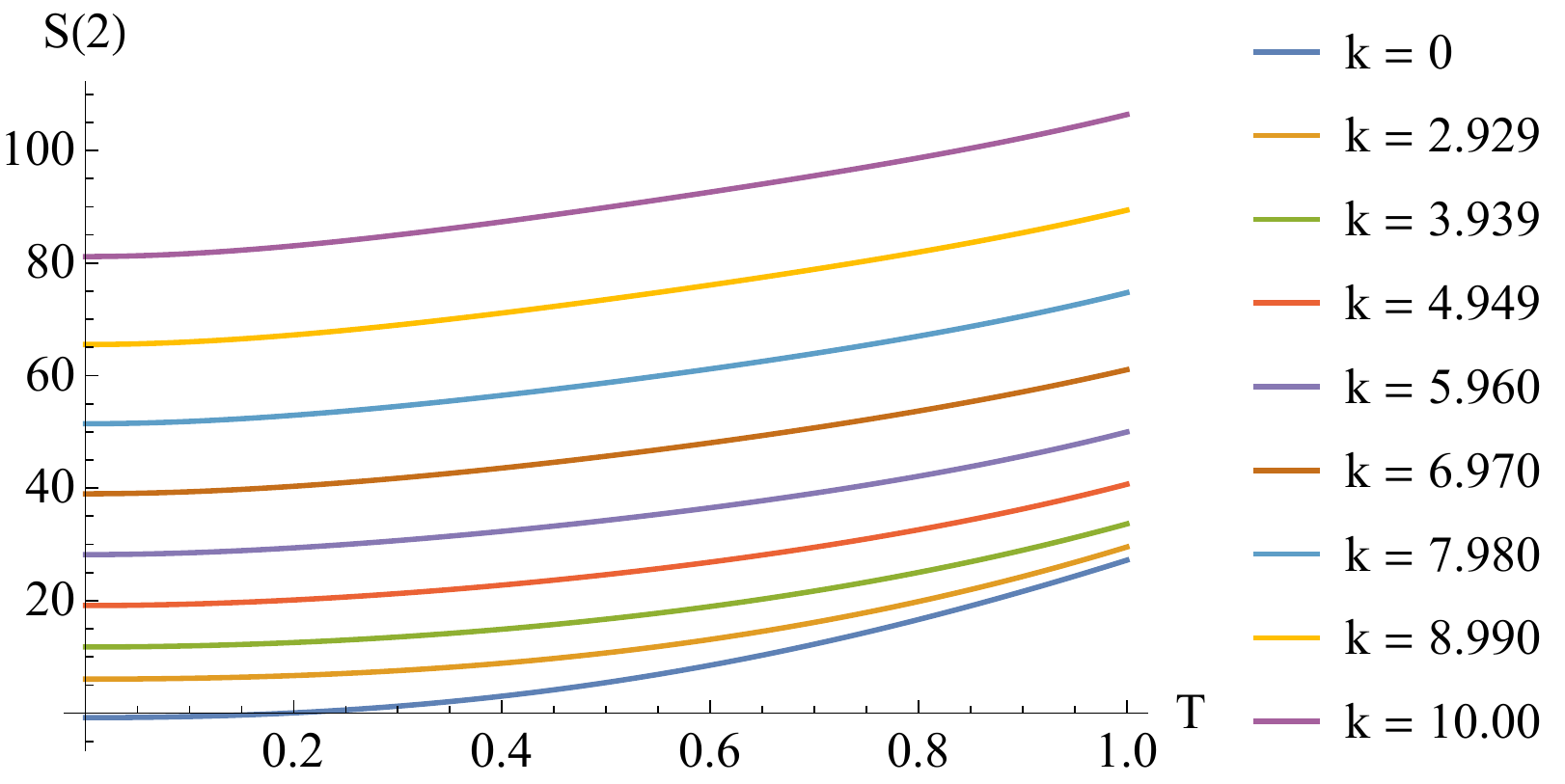}\\
	\includegraphics[width=0.75\textwidth]{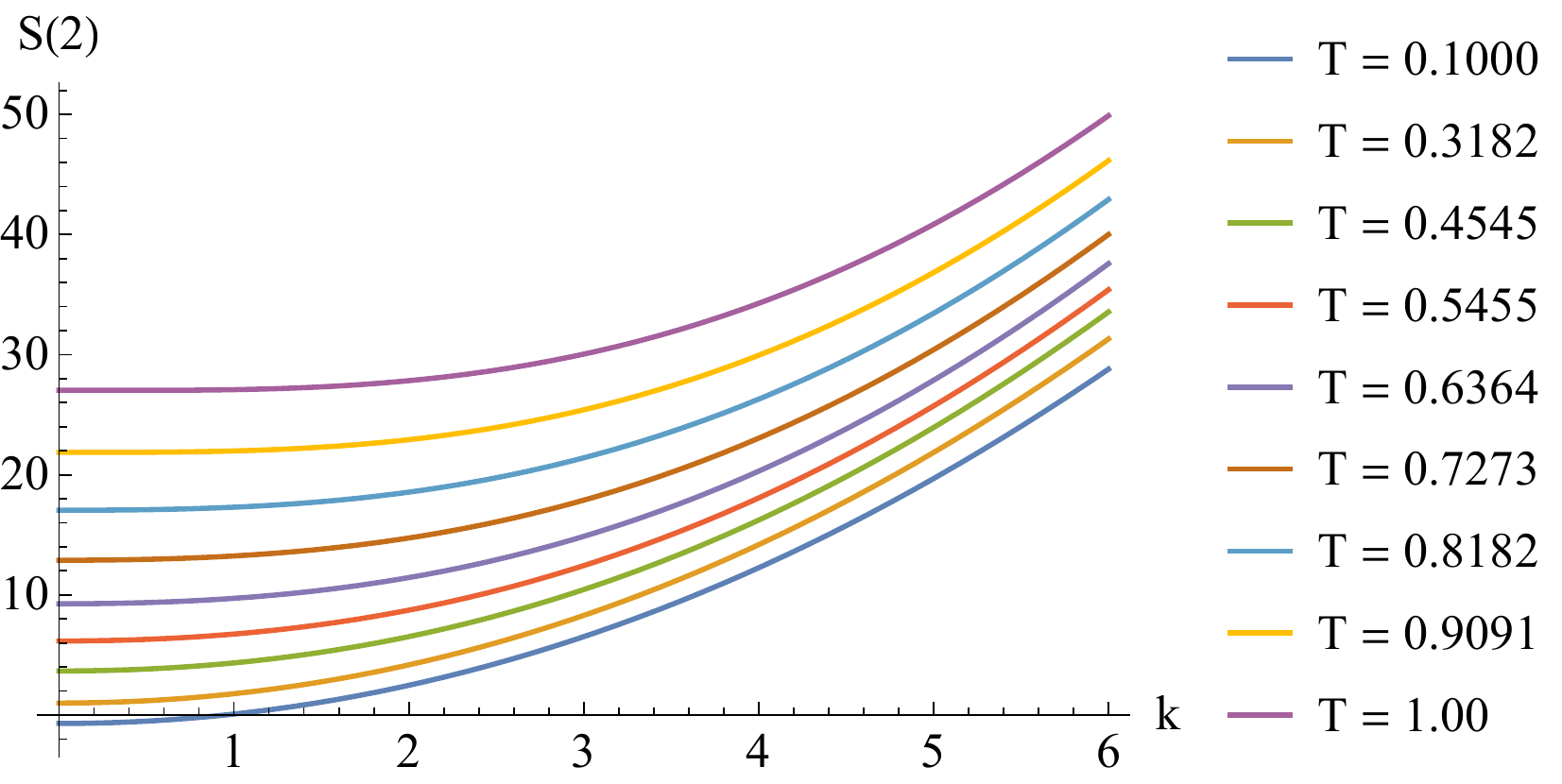}
	\caption{HEE at width $ w=2 $ vs $k$ (the upper plot) and $T$ (the lower plot).  For other values of $ w $ we have qualitatively similar behaviors.}
	\label{fig:heevsandt}
\end{figure}

Next, we show the HEE behavior with $ k $ and $ T $ at width $w=2$ respectively in Fig. \ref{fig:heevsandt}. Apparently, the HEE monotonically increases with $k$ and $T$, regardless of the values of $k, T$ and the subregion size. Notice also that HEE is almost flat in small $k$ region. This can be understood by the small $k$ expansion of the $\frac{\partial S}{\partial k}$.
\begin{equation}\label{eq:dsdkf1}
	\frac{\partial S}{\partial k} = \int_0^{w}
	\frac{16 k^3}{\left(12 r_h^2+1\right) \sqrt{1-\frac{4 z'(x)^2}{4 \left(z^3-1\right) r_h^2-z^4+z^3}}}
	\left[
		\frac{1}{z^2} + \frac{p(z) z'(x)^2}{1-z}
		\right]
	dx > 0 + \mathcal O(k^4),
\end{equation}
where $p(z)\equiv \frac{4 \left(3 z^3+2 z^2+2 z+2\right) r_h^2-3 z^3}{\left(z^4-4 z \left(z^2+z+1\right) r_h^2\right){}^2} >0$. Consequently, $\frac{\partial S}{\partial k} \sim k^3$ for small $k$, that explains the flat behavior of HEE.

Next, we study two different kinds of mixed state entanglement: MI and EoP.

\subsection{Mutual Information}\label{subsec:mi}

Compared with the monotonic dependence of HEE on system parameters, we find that MI presents more abundant phenomena. Moreover, the dependence of MI with $k$ is different from that of MI with $T$. Thus, we explore MI vs $k$ and MI vs $T$ respectively.

\subsubsection{Mutual Information vs $k$}

The MI of small configuration is different from that of large configuration, and the mechanisms behind these phenomena are also different. Therefore, we discuss the behavior of MI in small and large configurations respectively.

\begin{enumerate}

	\item Small Configurations

	      For small configurations, HEE and MI are mainly determined by the asymptotic AdS geometry. We may analytically explore the MI vs $k$.

	      We find that MI decreases monotonically with $k$ (see Fig. \ref{fig:ivsk4t}). According to the MI definition \eqref{eq:eps}, the monotonic behavior of $I$ corresponds to $\partial_k I = \partial_k S(A) + \partial_k S(B) - \partial_k S(B) - \partial_k S(A\cup B\cup C) < 0$. In fact, the $\partial_k I$ mainly depends on the $\partial_k S(A\cup B\cup C)$ because minimum surface of $A\cup B \cup C$ probes deeper into the bulk, and hence are more affected than the other three quantities by the deviation from the AdS geometry. Therefore, we can conclude that $\partial_k I<0$ following $\partial_k S>0$ from \eqref{eq:dsdkf1}.

	      \begin{figure}
		      \centering
		      \includegraphics[width=0.55\textwidth]{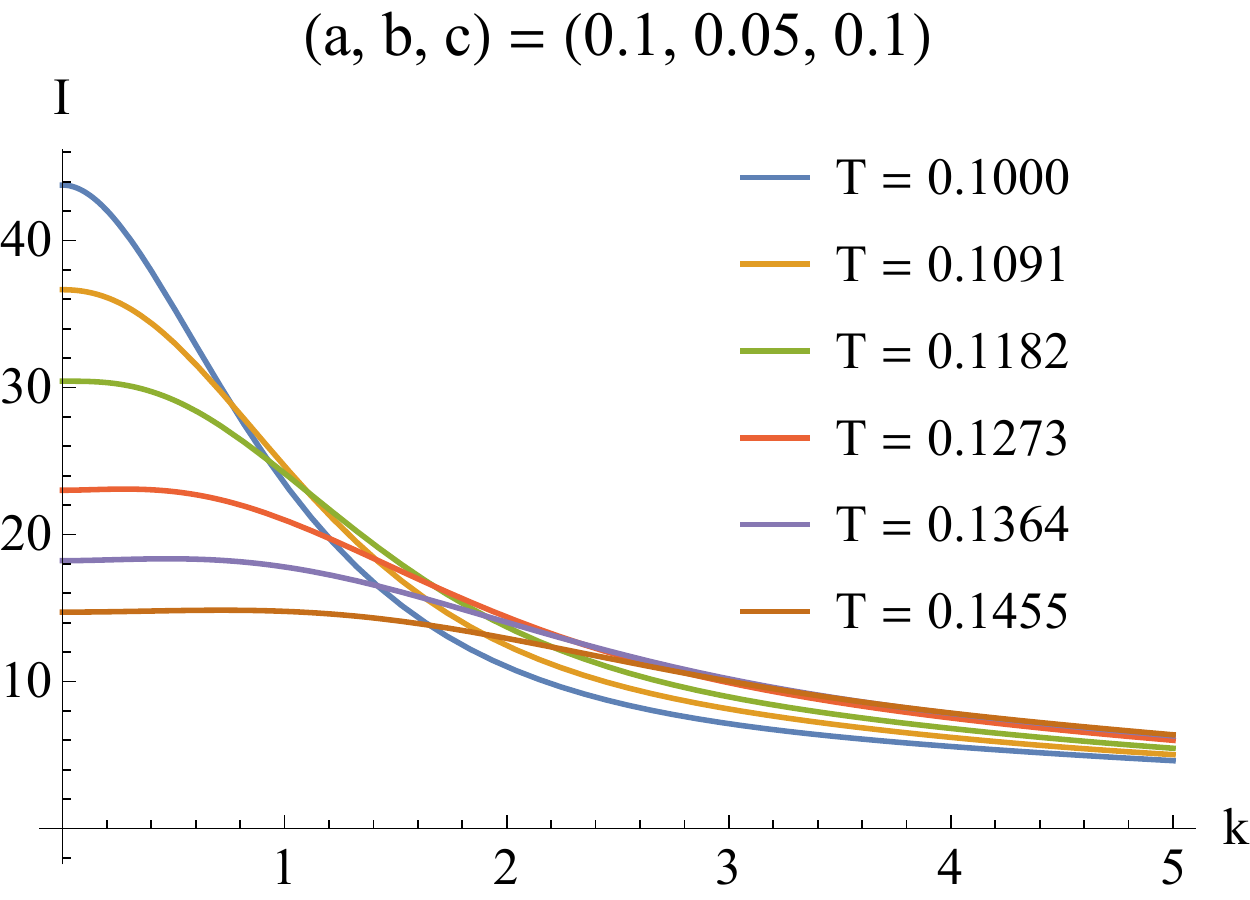}
		      \caption{The plot of $I$ vs $k$ at different temperatures specified by the plot legends. The small configuration is $(a,b,c) = (0.1,0.05,0.1)$.}
		      \label{fig:ivsk4t}
	      \end{figure}

	      Another significant phenomenon is that MI almost does not change with $k$ when $k$ is small, which can be understood by $\partial_k S\sim k^3$ from \eqref{eq:dsdkf1}.

	\item Large configuration

		  The minimum surfaces of a large configuration will stretch deep into the bulk, so it can not be understood analytically by the geometry deviation near the boundary. We use numerical method to explore MI vs $k$ for large configurations. 

	      When the configuration increases, we find that MI monotonically decreases with $k$ at the beginning. But when $a,\,c$ increases, we find that the MI becomes non-monotonic with $k$. No matter how large $a,\,c$ becomes, MI always monotonically decreases with $k$ in the small $k$ region. These phenomena are shown in Fig. \ref{fig:ivsk4ac}.

	      The universal monotonic behavior for small $k$ can also be understood from \eqref{eq:dsdkf1}. For large configurations, the MI will be dominated by $S(A), \, S(C),\,S(A\cup B\cup C)$ since they approach the horizon (see the second term of \eqref{eq:dsdkf1}). These three quantities are mainly contributed by the thermal entropy $s$. Therefore, the $\partial_k I$ will be determined by $-\partial_k s(B) <0$. As a result, we have $\partial_k I <0$ again.

	      \begin{figure}
		      \centering
		      \includegraphics[width = 0.55\textwidth]{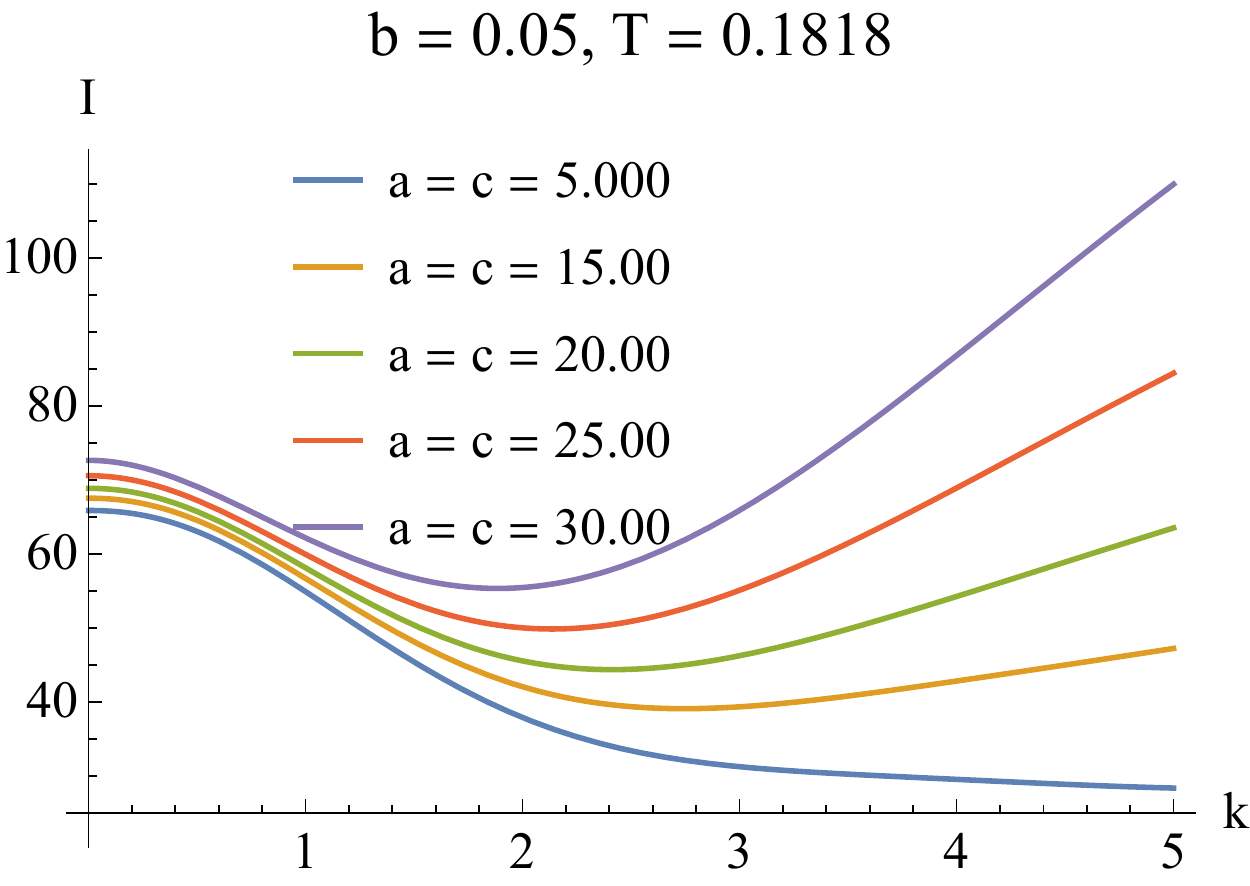}
		      \caption{The plot of $I$ vs $k$ at different configurations specified by the plot legends, at $b=0.05,\, T = 0.1818$. The behavior of MI at other temperatures is qualitatively the same.}
		      \label{fig:ivsk4ac}
	      \end{figure}

\end{enumerate}

Next we explore the relationship between MI and temperature.

\subsubsection{Mutual Information vs $T$}

\begin{figure}
	\centering
	\includegraphics[width = 0.55 \textwidth]{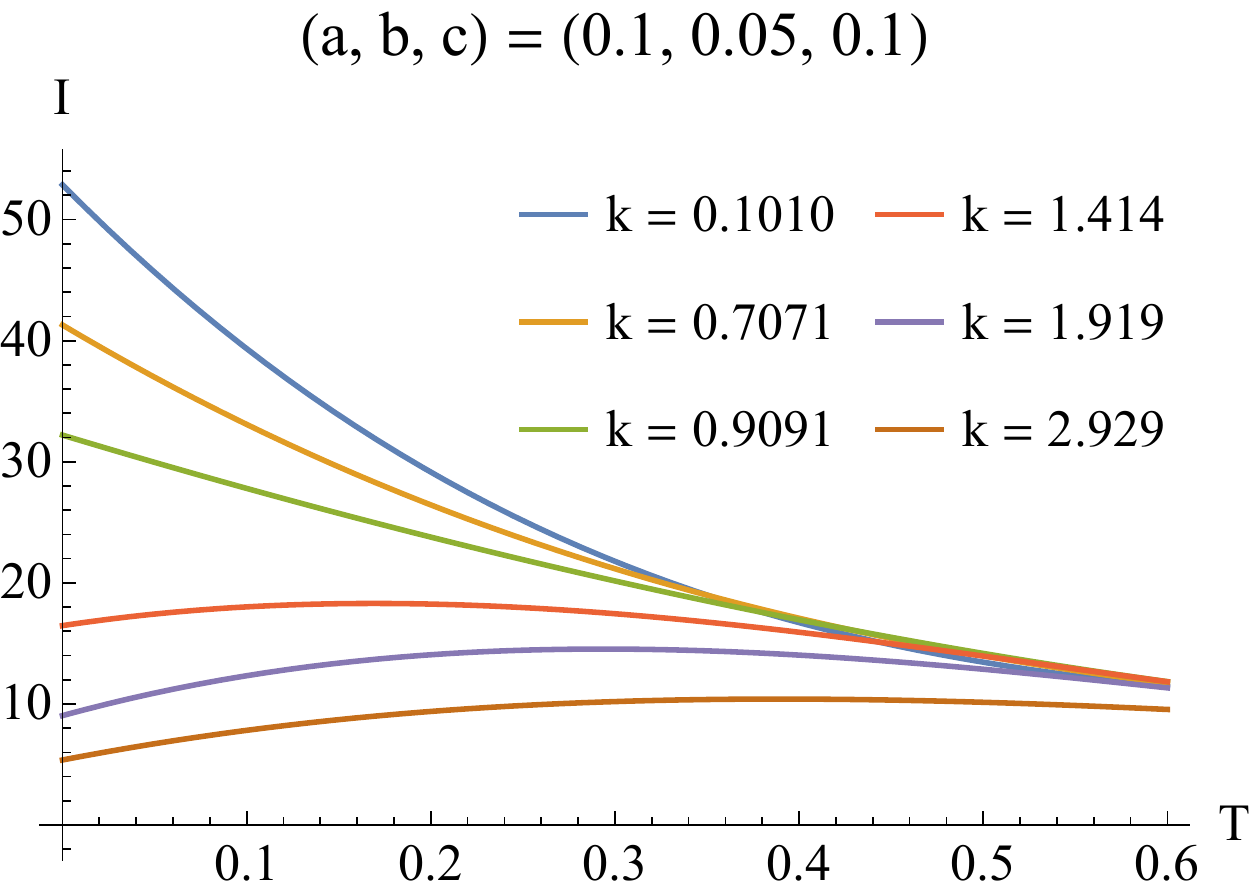}
	\caption{The plot of $I$ vs $T$ at different temperatures specified by the plot legends. The small configuration is $(a,b,c) = (0.1,0.05,0.1)$.}
	\label{fig:ivst4k}
\end{figure}

\begin{figure}
	\centering
	\includegraphics[width = 0.55 \textwidth]{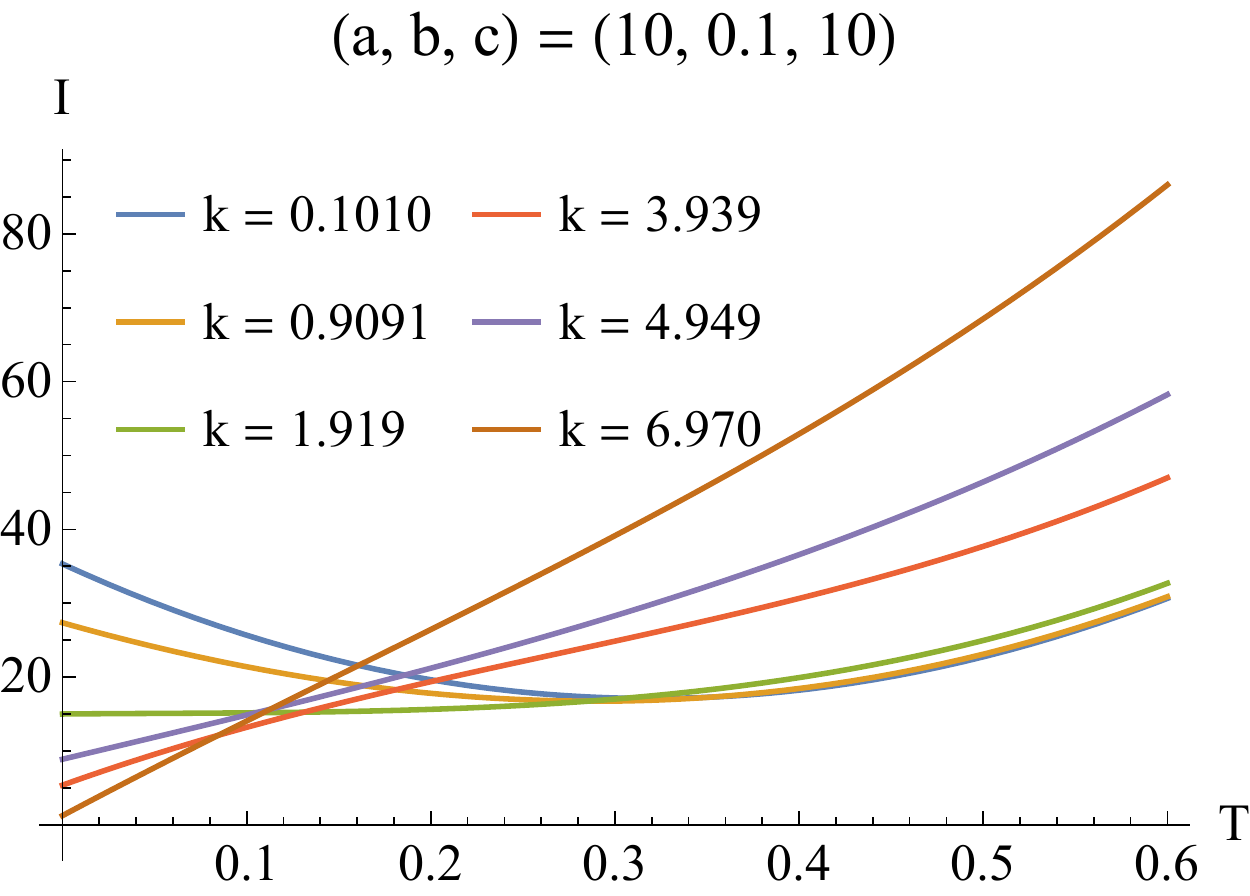}
	\caption{The plot of $I$ vs $T$ at different temperatures specified by the plot legends. The large configuration is $(a,b,c) = (10,0.05,10)$.}
	\label{fig:ivst4k2}
\end{figure}

The relationship between MI and temperature also relate to specific system parameters and configurations. Comparing Fig. \ref{fig:ivst4k} and Fig. \ref{fig:ivst4k2}, we see that when configuration increases, the behavior of MI with temperature becomes more abundant. Specifically, we find that $\partial_T I < 0$ when $k$ is small and $\partial_T I > 0$ when $k$ is relatively large, regardless of the configurations.

\begin{figure}
	\centering
	\includegraphics[width = 0.5\textwidth]{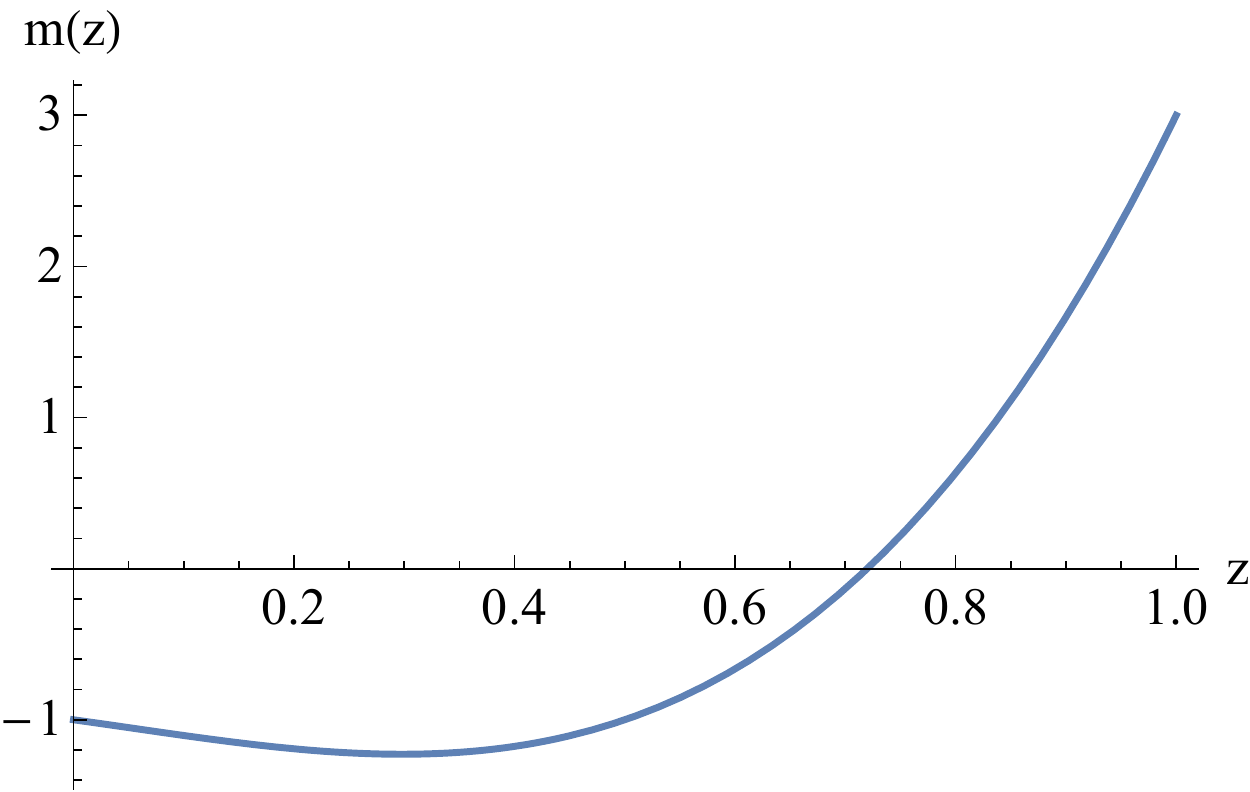}
	\caption{The plot of $m(z)$.}
	\label{fig:plotpm}
\end{figure}

First, we discuss the small $k$ limit. Taking the limit $k\to 0, \, T\to 0$, we find,
\begin{equation}\label{eq:smallkv1}
	\frac{\partial S}{\partial T} = \int_0^w \frac{1}{2 z^2 \sqrt{\frac{12 z'(x)^2}{(3 z-4) z^3+1}+1}} \left[1 - \frac{6 m(z) z'(x)^2}{(1-z)^3 (z (3 z+2)+1)^2}\right] dx
\end{equation}
where $m(z)\equiv z (2 z-1) (3 z+1)-1$. The sign of $\partial_T S$ depends on $m(z)$. From the plot of $m(z)$ in Fig. \ref{fig:plotpm}, we can deduce that $\partial_T S > 0$ for small configuration since the minimum surface resides only in the small-$z$ region. Following the arguments for the MI behavior for small configurations in the previous section, we see that $\partial_T I < 0$ for small configurations.

For large configurations, the minimum surface approaches the horizon and goes to the region with positive $m(z)$. At this time, the sign of $\partial_T S$ is not transparent from \eqref{eq:smallkv1}. However, the temperature behavior of the HEE as well as the MI, depends on the thermal entropy density since the HEE will be mainly contributed by the thermal entropy $s$. Therefore, we still have $\partial_T S>0$. Following the arguments for large configurations in the previous subsection, we have $\partial_T I < 0$ again.

Next, the study the large $k$ limit. The small $k$-limit analysis technique used in the previous section does not apply to the analysis of large k-limit. Because the radius of the horizon $r_h$ becomes large for large $k$. As a result, the minimum surface of any finite subregion approaches the horizon, leading to a vanishing MI. As can also be seen in Fig. \ref{fig:ivst4k} and Fig. \ref{fig:ivst4k2}, MI in the low-temperature region tends to vanish when $k$ is large enough. Therefore, the monotonous increase of MI with temperature in low temperature region occurs when MI is about to vanishes. At this time, the minimum surface is neither near the boundary nor near the black hole horizon. Therefore, the analysis techniques of large and small configuration limits are also not applicable. At this stage, we can only numerically address this phenomenon. We plan to explore an analytical understanding in the near future.

Next we examine the behavior of the EoP.

\subsection{Entanglement of Purification}\label{subsec:eop}

The EoP for a symmetric configuration equals the area of the vertical line connecting the tops of the minimum surfaces (see Fig. \ref{fig:midemo}). For simplicity, we focus on the EoP of symmetric configurations in this paper. Most of the previous studies on EoP only considered symmetric configurations \cite{Takayanagi:2017knl,Nguyen:2017yqw,Umemoto:2018jpc,Yang:2018gfq,Hirai:2018jwy,Tamaoka:2018ned,Espindola:2018ozt,Agon:2018lwq,Guo:2019azy,Ghodrati:2019hnn}. For asymmetric configurations ($a\neq c$), one needs to search a two-dimensional parameter space to find the EoP, which is a hard task \cite{Liu:2019qje}. 

\begin{figure}
	\centering
	\includegraphics[width = 0.45\textwidth]{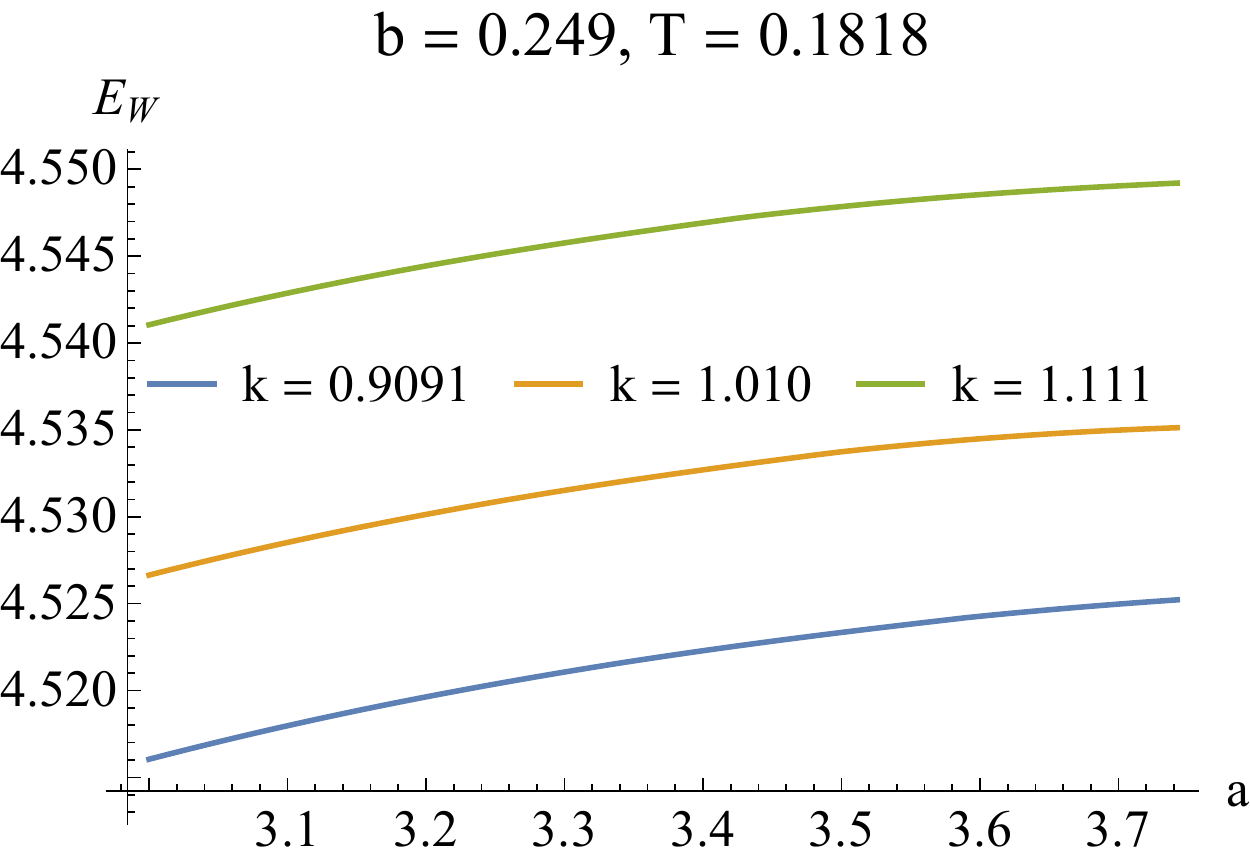}
	\includegraphics[width = 0.45\textwidth]{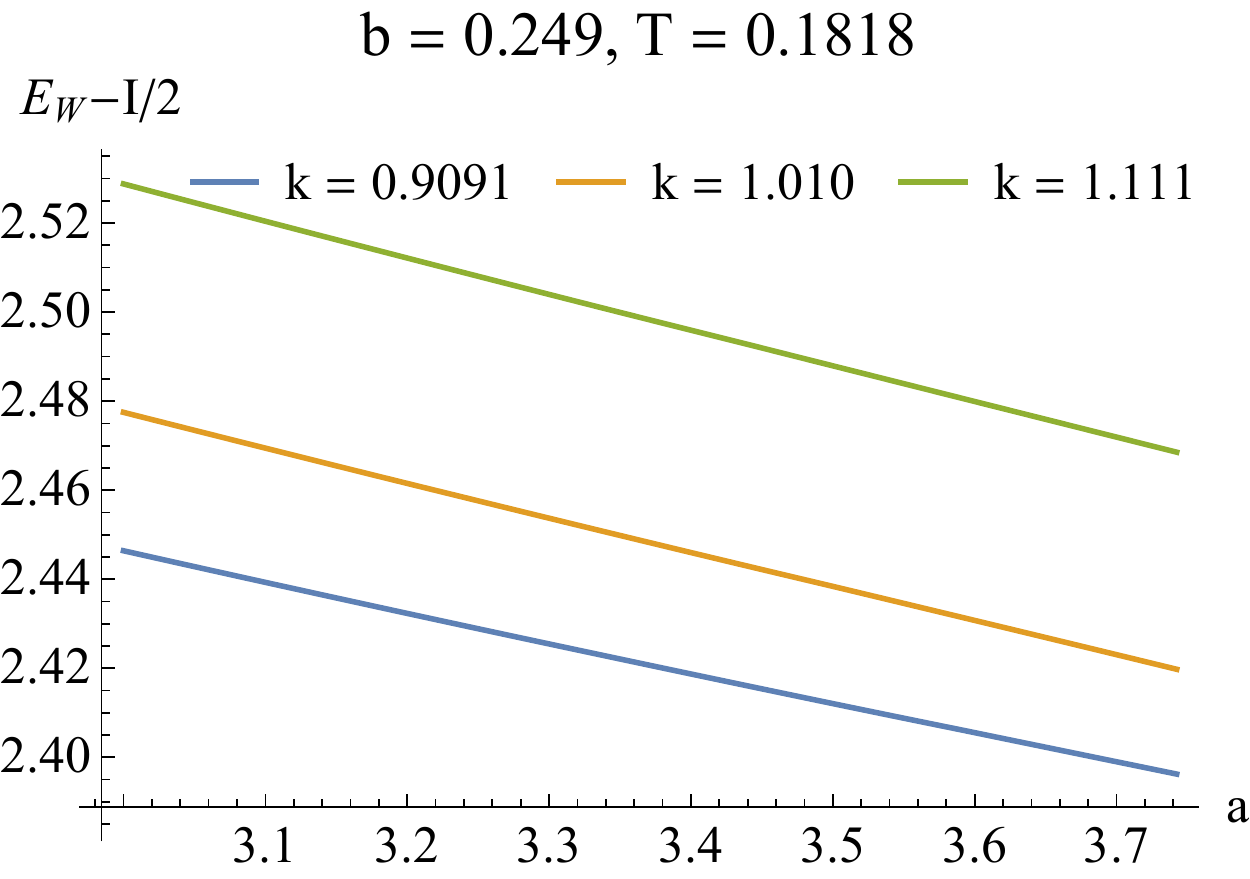}
	\caption{The first plot: EoP vs $a, c$. The temperature is fixed as $t = 0.1818$, and the separation $b$ is fixed as $0.249$. Different curves in the diagram represent different $k$ values  specified by the plot legends. The second plot: $E_W - I/2$ vs $a$. The second plot shows that $E_W$ is always greater than half of the mutual information $I/2$.}
	\label{fig:eopvsa4bkt}
\end{figure}

\begin{figure}
	\centering
	\includegraphics[width = 0.45\textwidth]{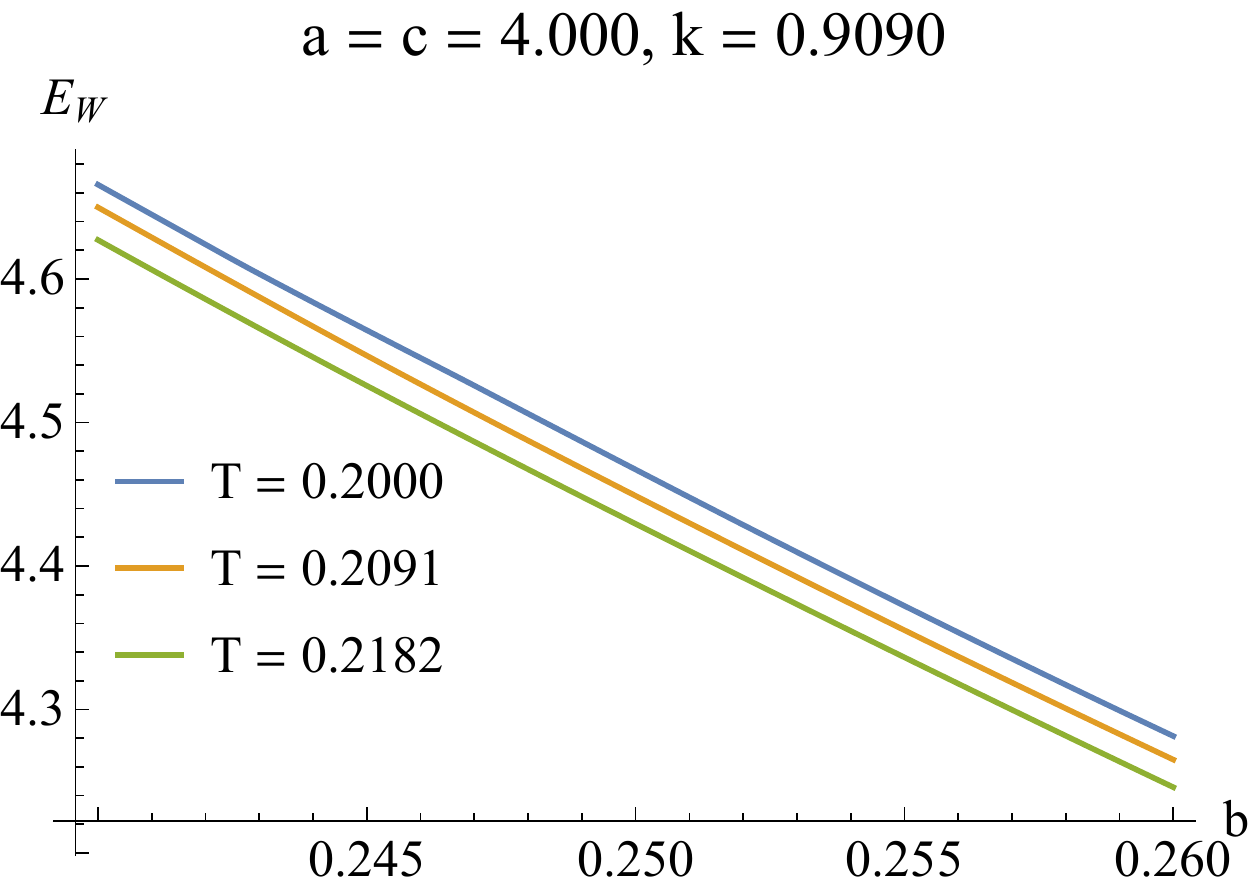}
	\includegraphics[width = 0.45\textwidth]{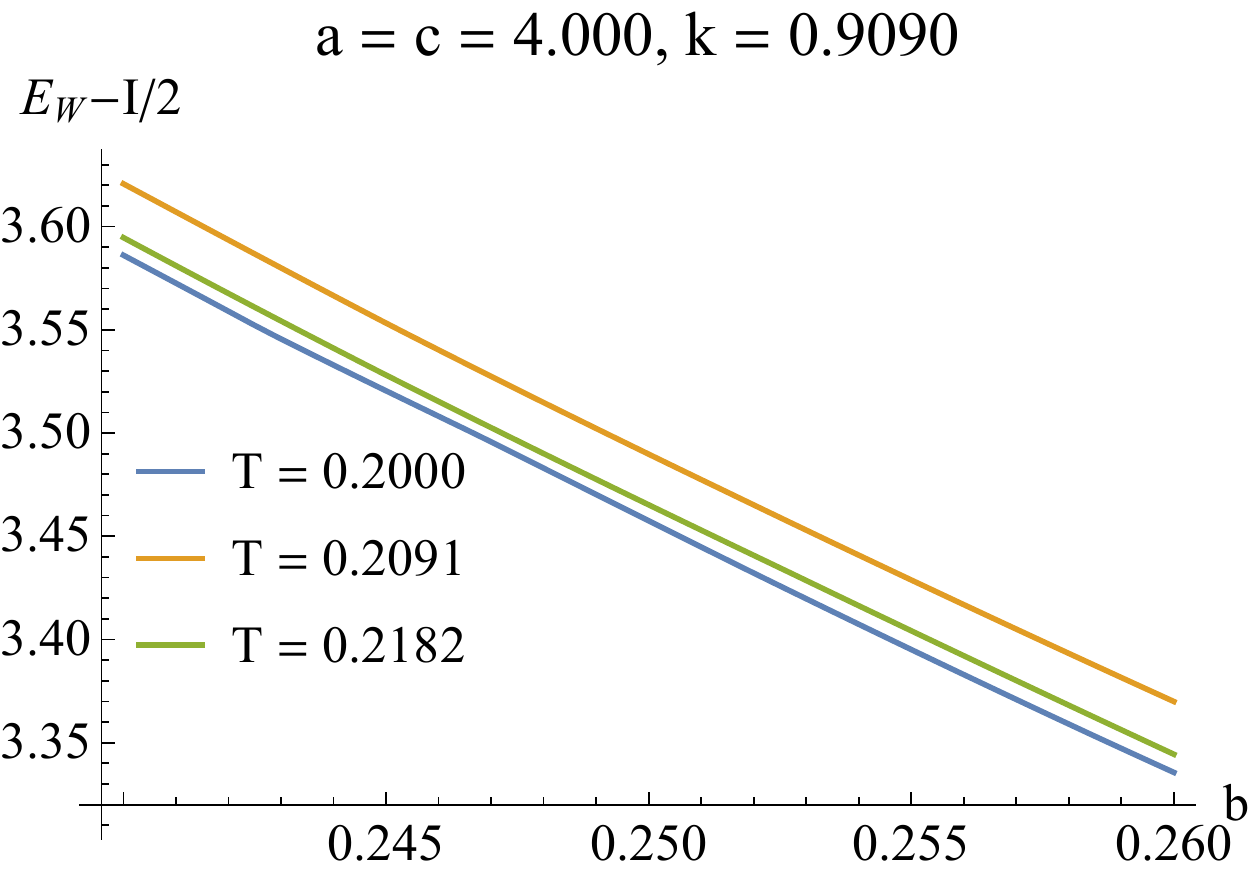}
	\caption{The first plot: EoP vs $b$ at $k=0.909$ and $a=c=4$. The second plot: $E_W - I/2$ vs $b$. The second plot shows that $E_W$ is always greater than half of the mutual information $I/2$. Each curve in both plots corresponds to $k$ specified by the plot legends.}
	\label{fig:eopvsb4akt}
\end{figure}

We show the relationship of EoP with $a, c$ in Fig. \ref{fig:eopvsa4bkt}, and EoP with $b$ in Fig. \ref{fig:eopvsb4akt}. EoP increases with the increase of $a, c$ and decreases with the increase of $b$. These phenomena verify the inequality $E_{W}\left(\rho_{A(B C)}\right) \geqslant E_{W}\left(\rho_{A B}\right)$ satisfied by EoP \cite{Takayanagi:2017knl}. In addition, from the right plots in Fig. \ref{fig:eopvsa4bkt} and Fig. \ref{fig:eopvsb4akt} we see that EoP is always larger than $I/2$. This verifies another important inequality $E_{W}\left(\rho_{A C}\right) \geqslant \frac{I(A;C)}{2}$ that EoP satisfies \cite{Takayanagi:2017knl}.

Next, we study EoP vs $k$ and EoP vs $T$.

\subsubsection{Entanglement of Purification vs $k$}

First, EoP monotonically increases with $k$, regardless of the specific system parameters and configurations. When $k$ reaches the critical $k_c$, EoP vanishes because MI vanishes. We show these phenomena in Fig. \ref{fig:ploteopvsk}. The monotonic behavior can be seen from,
\begin{equation}\label{eq:dewdk}
	\frac{\partial E_W}{\partial k} = \int_{z_{*1}}^{z_{*2}} \frac{8 k^3 r_h \left(n(z)+2 \pi  T z^3\right) \sqrt{\frac{r_h}{(1-z) n(z)}}}{z^2 n(z) \left(12 r_h^4+r_h^2+6 k^4\right)} dz > 0,
\end{equation}
where $n(z) \equiv \left(-3 z^3+z^2+z+1\right) r_h+4 \pi  T z^3$, the $z_{*1}$ and $z_{*2}$ are the tops of the minimum surfaces (see Fig. \ref{fig:midemo}). Therefore, we conclude that $E_W$ monotonically increases with $k$. This fact shows that the bipartite entanglement of the dual field theory of the axion model always increases with the $k$.

\subsubsection{Entanglement of Purification vs $T$}

\begin{figure}
	\centering
	\includegraphics[width= 0.55 \textwidth]{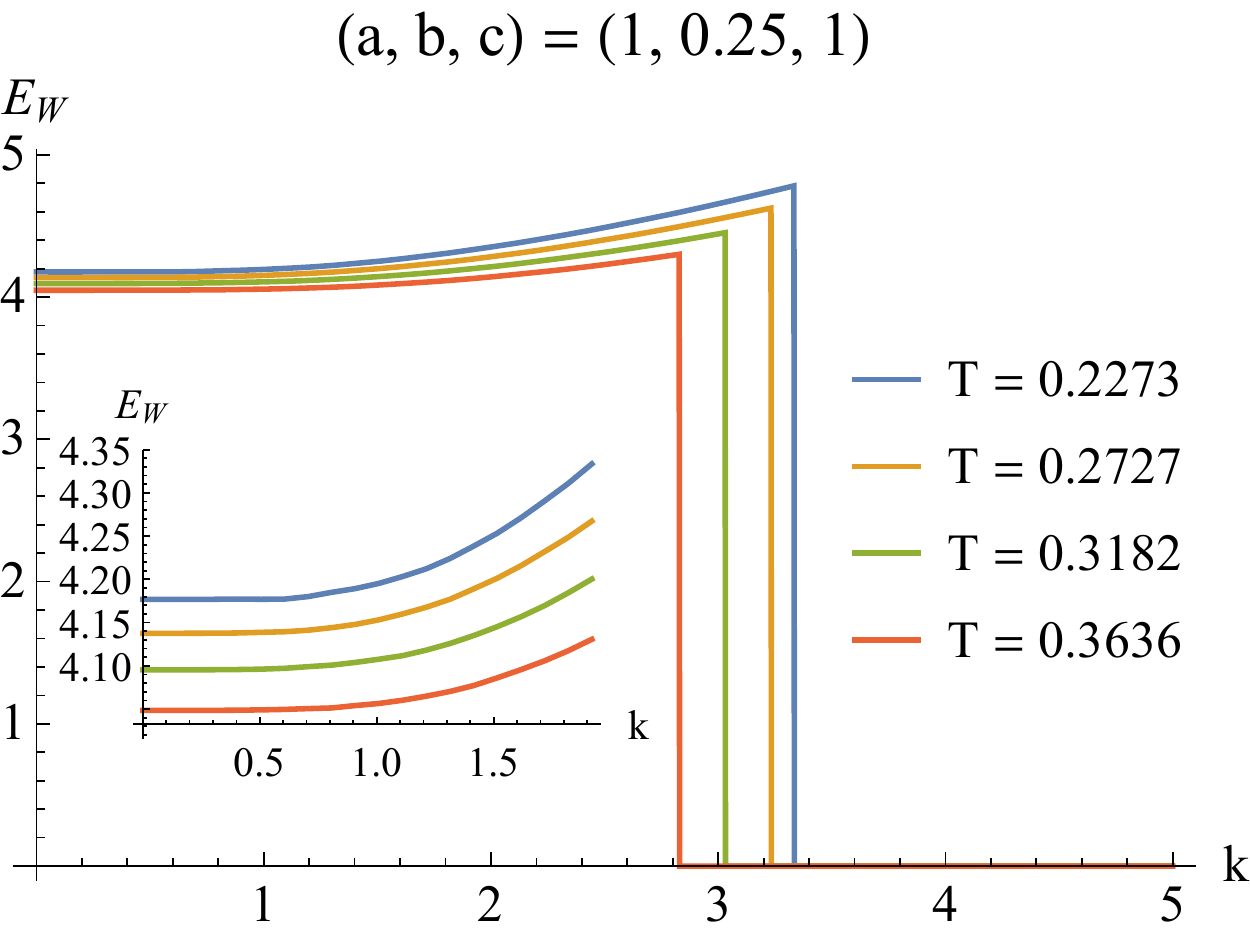}
	\caption{The relation between EoP and $k$. The reason for the sudden drop of EOP to 0 is that the MI under this configuration is actually 0. That is to say, this point corresponds to the critical point of disentangling transition.}
	\label{fig:ploteopvsk}
\end{figure}

For small configurations, we find that EoP increases with $T$, regardless of the values of $k, T$. This can be deduced from,
\begin{equation}\label{eq:dewdt}
	\frac{\partial E_W}{\partial T} = \int_{z_{*1}}^{z_{*2}} \frac{4 \pi  r_h \sqrt{\frac{r_h}{(1-z) n(z)}} \left(4 r_h^3 n(z)-z^3 \left(r_h^2+4 k^4\right)\right)}{z^2 n(z) \left(12 r_h^4+r_h^2+6 k^4\right)} dz,
\end{equation}
where the sign of $\frac{\partial E_W}{\partial T}$ depends on $4 r_h^3 n(z)-z^3 \left(r_h^2+4 k^4\right)$. This term is always positive for small configurations, which results in a monotonically increasing behavior with $T$. Therefore, we have $E_W'(T) > 0$ for small configurations. We also plot this phenomenon in Fig. \ref{fig:ploteopvsk}.

\begin{figure}
	\centering
	\includegraphics[width= 0.55 \textwidth]{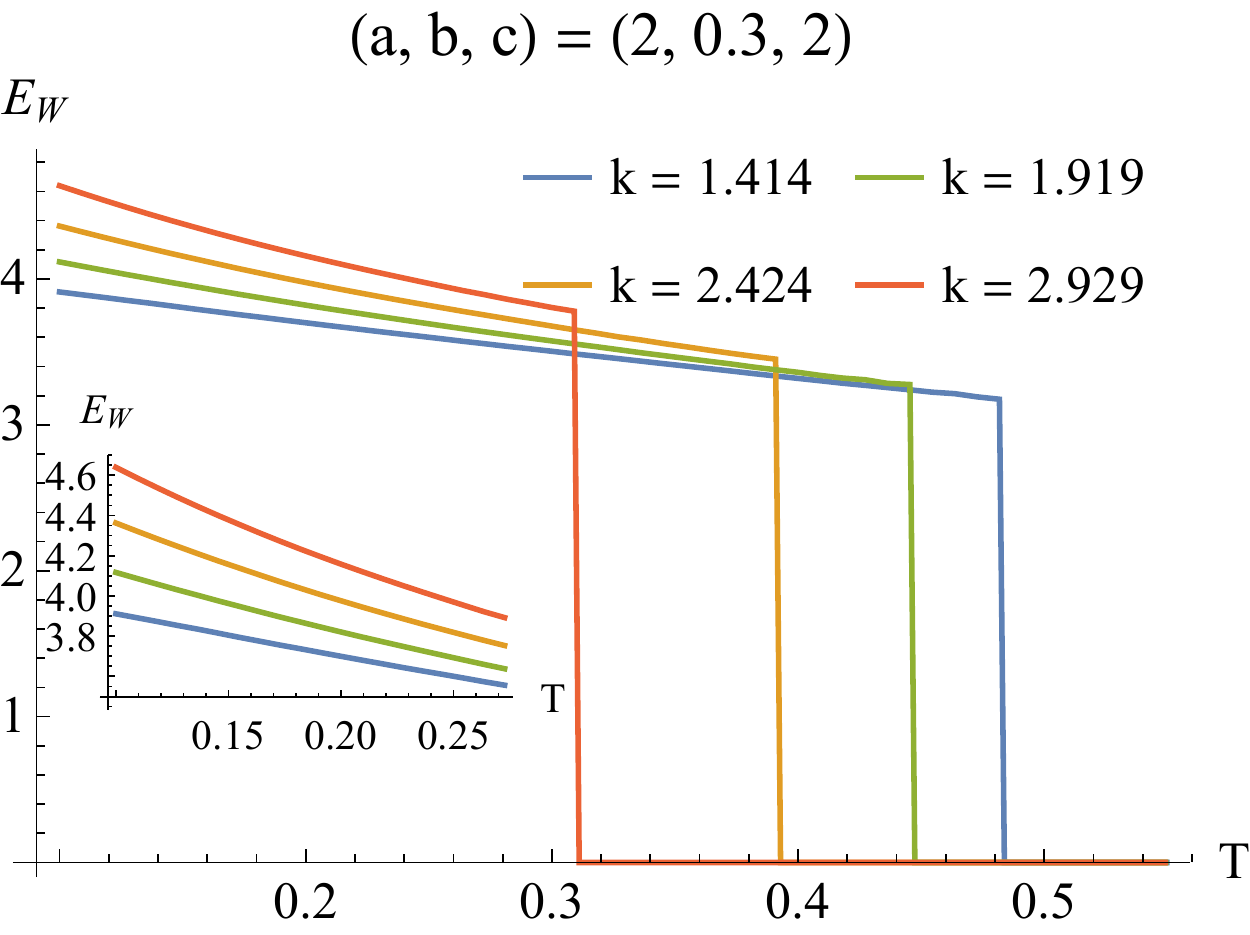}
	\caption{The relation between EoP and $T$. The reason for the sudden drop of EOP to 0 is that the MI under this configuration is actually 0. That is to say, this point corresponds to the critical point of disentangling transition.}
	\label{fig:ploteopvstv1}
\end{figure}

However, for large configurations, a monotonic behavior of $E_W$ with $T$ is not guaranteed. We have also analyzed $\partial_T E_W$ in small $k, T$ limit and in large configuration limit, but we did not find any monotonic behavior. The relation between $E_W$ and $T$ depends on specific system parameters and configurations. Therefore, We resort to numerics to explore $E_W$ vs $T$. We plot $E_W$ vs $T$ in Fig. \ref{fig:ploteopvstv1}, in which the $E_W$ decreases with $T$ before it vanishes. We remind that $E_W$ can exhibit non-monotonic behavior for other system parameters $k, T$. Meanwhile, we notice that EoP will vanish as MI vanishes when $T$ is large. This is because $r_h$ becomes large for high temperature, and the minimum surface of any finite subregion approaches the horizon. This situation of large $T$ limit is the same as that of the large $k$ limit, where we have vanishing MI. The vanishing $E_W$ at large temperatures means that the thermal effects will break the entanglement between separate subregions.

\subsection{Comparison of three entanglement measures}

We have studied the behavior of HEE, MI and EoP on axion model, and found that they exhibit very distinct behaviors. First, HEE increases with $k$ and $T$ monotonically, which can be understood analytically in certain limits. However, the monotonic behavior of HEE does not mean that the entanglement of the system increases with $k, T$, because the thermal entropy can bury the quantum entanglement \cite{Fischler:2012uv}. Especially, HEE is actually dictated by thermal entropy for large $k$ limit and large $T$ limit.

The MI typically exhibits non-monotonic behaviors with $k$ and $T$. Nevertheless, the MI decreases with $k$ monotonically when $k$ is small. This fact can be proved analytically for small configurations and large configurations. For low temperatures, the MI always decreases with $T$ for small $k$; while for large enough $k$, we observed a universal increasing behavior with $T$. These phenomena suggests that MI captures distinct entanglement structure from the HEE. However, MI may still be dictated by the thermal entropy due to its dependence on HEE, especially for large configurations.

The EoP monotonically increases with $k$ before it vanishes, which can be proved analytically. This unusual monotonic behavior implies that the EoP captures very different entanglement structures from the MI, since MI does not show any universal monotonic behavior. Moreover, the EoP monotonically decreases with $T$ for small configurations, this is another distinct property compared with MI, remind that the MI increases with $T$ for large enough $k$. More importantly, the EoP for large configurations cannot be analyzed by near horizon geometry, which we have adopted to analytically deduce the monotonic behavior for HEE and MI. The reason is that the EoP involves the bulk degrees of freedom even in the large configurations limit. That is to say, the EoP will never be dictated by the thermal effects. Therefore, the EoP may be a better mixed entanglement measure than the MI.

\section{Discussion}\label{sec:discuss}

In this paper, we have studied HEE, MI and EoP on holographic axion model, and found that they exhibit very different behaviors. Combined with numerical and analytical analysis, we found and proved their monotonic behavior in certain limits. Their differences show that they depict different aspects of the entanglement properties. Specifically, MI and EoP can cancel out the thermal effect compared with HEE, and hence exhibits more diverse phenomena. Moreover, EoP can be a better mixed state entanglement measure than the MI due to its independence from the thermal entropy. Next, we point out several topics worthy of further study.

First, the techniques in this paper can be directly applied to holographic models with analytical solutions, such as Gubser-Rocha model \cite{Gubser:2009qt}, massive gravity theory \cite{Cai:2014znn}, Gauss-Bonnet gravity theory \cite{Cai:2001dz}, and so on. For numerical background solutions, in certain limits, the techniques in this paper may still be applicable. In addition, the EoP of asymmetric configuration in this paper is worth studying, it would be interesting to test whether the monotonic behavior of $E_W$ with $k$ is universal. Finally, it is also desirable to study the behavior of HEE, MI and EoP during phase transitions.

\section*{Acknowledgments}
Peng Liu would like to thank Yun-Ha Zha for her kind encouragement during this work. This work is supported by the Natural Science Foundation of China under Grant No. 11805083, 11847055, 11905083.

\end{document}